\documentclass[5p, sort&compress]{elsarticle}

\usepackage{amsmath}
\usepackage{graphicx}

\providecommand{\e}[1]{\ensuremath{\times 10^{#1}}}

\begin{document}

\begin{frontmatter}

\title{The Effects of Brownian Motion On Particle Interactions with \\
Patchy Surfaces Bearing Nanoscale Features}

\author[cheEng]{Marina Bendersky}
\author[pse]{Maria M. Santore}
\author[cheEng]{Jeffrey M. Davis}

\address[cheEng]{Department of Chemical Engineering, University of Massachusetts, Amherst, MA 01003}
\address[pse]{Department of Polymer Science and Engineering, University of Massachusetts, Amherst, MA 01003}

\begin{abstract}

The effects of Brownian motion on particle interactions with heterogeneous collectors are evaluated by adding stochastic Brownian displacements to the particle trajectories and comparing those trajectories to those where Brownian motion is not included. We define adhesion thresholds as the average patch density on the collector required to adhere colloidal particles and P\'{e}clet numbers that quantify the relative importance of colloidal, shear and Brownian effects.

We show that Brownian motion has a negligible influence on particle trajectories over collectors patterned with nano-scale heterogeneity, the non-uniform distribution of which creates locally attractive and repulsive areas within the collector. High energy barriers in strong locally repulsive areas cannot be overcome by Brownian motion, such that particle deposition on patchy collectors is controlled by spatially varying DLVO interactions and not by Brownian motion. The overall adhesive behavior of the system remains unaffected by the introduction of Brownian motion effects in the simulations, and therefore, 
for particle sizes that are usually used in experiments of particle trajectories over nanoscale heterogeneous collectors, it is reasonable to neglect Brownian motion effects entirely.

\end{abstract}

\end{frontmatter}

\section{Introduction}

Interest in the study of particle deposition mechanisms on heterogeneous substrates \cite{AdamczykPatterned:2009, Adamczyk:2001} has been recently increasing, not only due to the critical role that particle-collector interactions play in current industrial applications \cite{KimZydney:2004, SandEtAl:2009}, but also due to the relevance of these studies to the development of emerging sensing \cite{biosensorsKatz:2001} and separation \cite{BiotechProducts:2007} technologies at the micro- and nano-scales. 
Recent studies of particles flowing over collectors patterned with nanoscale patches, which are orders of magnitude smaller than the depositing particles, revealed that small amounts of randomly-distributed, attractive patches induce particle deposition onto net-repulsive surfaces 
\cite{DuffadarDavisSantore:2009,DuffadarDavisFriction:2008,Kalasin:2008,SantoreKozlova:2006,SantoreKozlova:2007}.
Deposition rates for such systems are found to be larger than expected from a mean-field application of the classical DLVO theory \cite{BenderskyDavis:2011}. The discrepancies between 
empirical and predicted results are attributed to particle interactions with many patches, which
create locally attractive regions (``hot spots") \cite{DuffadarDavis:2007} on the collector as a result of their non-uniform distribution.   

Therefore, particle deposition on collectors patterned with nano-scale heterogeneity is found to be controlled not only by the total amount of attractive charge on the collector, but also, by 
the spacing between the particle-attracting patches. For surfaces densely covered by
patches (small patch-patch spacing), adhesion is rapid and transport-limited. For larger
patch-patch spacing, however, adhesion rates become slower and the amount of deposited
particles also decreases \cite{Kalasin:2008}.
The dependence of the adhesion capability of the collector on the patch-patch spacing, is thus better characterized by the system's adhesion threshold. The adhesion threshold is usually defined as the critical fraction of collector surface area covered with heterogeneity below which particles do not adhere on the collector. Alternatively, it can be defined as the critical patch-patch spacing above which silica particles do not deposit on the collector \cite{SantoreKozlova:2006,SantoreKozlova:2007}.        

In theoretical and computational studies \cite{DuffadarDavis:2007, DuffadarDavisFriction:2008, DuffadarDavisSantore:2009} of such systems, comprising flat and spherical electrostatically non-uniform collectors in which particles deposit from a flowing solution, Brownian motion effects have typically been neglected. The displacements of free particles due to Brownian motion, however, are often significant in time intervals characteristic of the imposed flow. 
Brownian motion can enable particle deposition on both homogeneous and heterogeneous collecting surfaces, even in the presence of an energy barrier \cite{AdamczykRSA:2005, Walz:1998}.

While adhesive dynamic simulations \cite{ModyKing:2007} showed that bond dissociation dynamics are not significantly influenced by Brownian forces and recent studies on particle interactions with rough collectors \cite{KempsBhattacharjee:2009} also indicated that Brownian motion effects are negligible, microchannel flow experiments of particle deposition \cite{UnniYang:2005} suggest that,   
for low-energy barrier systems, Brownian motion can indeed increase particles' tendency to adhere on heterogeneous collectors.

Brownian motion effects are modeled in this work with the ``Brownian displacements" approach, which consists on introducing stochastic Brownian displacements in Langevin-type particle trajectory equations \cite{ErmakMcCammon:1978, UnniYang:2005, KempsBhattacharjee:2009, AnsellDickinson:1986, Warszynski:2000}. 
Previous work focused on the study of the dynamics and aggregate formation properties of many-particle systems \cite{ErmakMcCammon:1978, AnsellDickinson:1986},
on particle deposition on homogeneous collectors \cite{KempsBhattacharjee:2009} or on particle behavior in parallel-plate microchannel flow \cite{UnniYang:2005}. 
In this study, Brownian motion effects are investigated for the case of a spherical particle interacting with collectors patterned with nano-scale heterogeneity. 

The focus of the present work is, therefore, to elucidate the specific effects of Brownian motion on particle interactions with collectors patterned with flat circular patches or with cylindrical pillars that protrude a few nanometers from the collector. 
The model systems resemble experimental set-ups used to study deposition mechanisms of colloidal particles, such as latex spheres that deposit on substrates covered with disk-shaped or spherical adsorption sites \cite{AdamczykRSA:2005}, or silica particles that adhere from flowing solution on patterned planar collectors \cite{SantoreKozlova:2006, SantoreKozlova:2007}.  
In experiments performed by Santore and Kozlova \cite{SantoreKozlova:2006,SantoreKozlova:2007}, uniformly and negatively charged 460-nm diameter silica particles flowed in solution over planar negatively charged silica surfaces patterned with positively charged round patches. Controlled amounts of the cationic polyelectrolyte pDMAEMA (poly-dimethyl-aminoethyl methacrylate) are deposited on acid-washed microscope slides to form round 11-nm diameter patches, which protrude from the surface only about 1 nm, remaining stable against aggregation and clustering. 
Due to the shear flow (at rates varying from 10 to 50 sec$^{-1}$), the patches are randomly and irreversibly deposited on the surface, and varying patch densities are obtained by allowing the pDMAEMA solution to flow over the negatively charged surface for different periods of time. 

Detection of adsorbed polyelectrolyte layers through particle deposition experiments \cite{AdamczykPElayers:2006,AdamczykRSA:2005} can lead to the development of pattern recognition technologies for sensing or encrypting purposes. 
As well, the efficiency of industrial processes in the fields of filtration, chromatography, and separation of proteins, viruses and bacteria, can be improved by using interfaces patterned with binding agents, such as polyelectrolytes \cite{AdamczykRSA:2005}, that can target desired pollutants with greater specificity.

The current study presents computations of particle trajectories as they translate and rotate in shear flow over heterogeneous collectors, being subject not only to DLVO interactions with the nano-scale collector heterogeneity but also to Brownian motion effects.
Electrostatic double layer (EDL) and van der Waals (vdW) forces and energies (DLVO interactions) are computed by implementing the grid-surface integration (GSI) technique \cite{DuffadarDavis:2007, BenderskyDavis:2011}. The shear-induced force and torque acting on the particle as it translates over a collector and the DLVO forces that result from particle interactions with it, are incorporated in a mobility matrix \cite{MobMatrixHammer:1996} formulation of the hydrodynamics problem to yield the particle's velocities.
Brownian motion effects are modeled by updating the particle's position with stochastic Brownian displacements \cite{KempsBhattacharjee:2009}.
The relative importance of shear, colloidal, and Brownian effects is evaluated through the computation of appropriate P\'{e}clet numbers, which are unique for this heterogeneous system.

\section{Methods}
\label{sec:Methods}

\begin{figure}[h!]
\centering
\includegraphics[width = 3.3in]{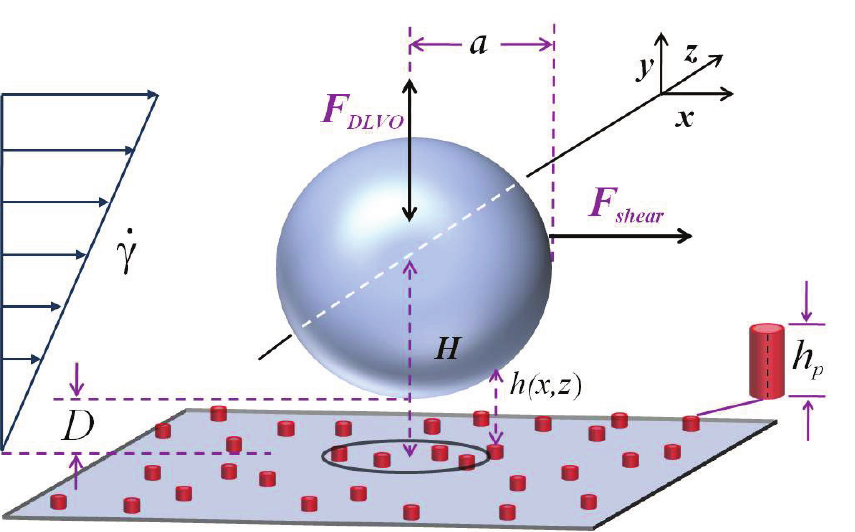}
\caption{Schematic diagram of a spherical particle of radius $a$ interacting with a heterogeneous surface patterned with randomly distributed nano-pillars of height $h_\text{p}$. The local separation distance between the sphere and the collector element vertically below it is $h(x,z)$ and $D$ is the minimum separation distance between the sphere and the plane defined by the top of the pillars. Shear-induced and colloidal forces act in the directions parallel and normal to the collector, respectively.}
\label{fig:schem}
\end{figure}

The model system, schematically illustrated in Fig. \ref{fig:schem}, depicts a colloidal particle of radius $a$ translating in shear flow parallel to a chemically and topographically heterogeneous collector. 
The collector heterogeneities are located at randomly chosen positions, 
to satisfy Poisson statistics as a first approximation of distributions of patches attained in experiments \cite{DuffadarDavisSantore:2009}. 
The fraction of surface area of the collector covered with heterogeneities is denoted by $\Theta$. The heterogeneities consist on either flat patches (with pillar height $h_\text{p} =0$ nm) or cylindrical pillars ($h_\text{p} = 2$ nm) that are charged with an electrostatic surface potential of $\psi_\text{patch,\,pillar} = 50$ mV.
The surface potential of the bare collector and of the flowing particle are $\psi_\text{sphere} = \psi_\text{collector} = -25\,$mV.

Brownian effects are modeled by introducing Brownian displacements \cite{KempsBhattacharjee:2009, bacteriaTransportNelsonGinn:2005, ElimelechEtAlbook:1995} in the parallel and normal directions to the collector surface. 
This approach parallels the Brownian Dynamics method \cite{ErmakMcCammon:1978, AnsellDickinson:1986, Warszynski:2000} and yields a Langevin-type equation for the particle trajectory. In the present study, the general form of such equation reads
\begin{equation}
\mathbf{x}(t) = \mathbf{x}^o(t) + \mathbf{U} \Delta t + \mathbf{R_\text{Br}}\,,
\label{eqn:LangevinEqn}
\end{equation}
where $\mathbf{x}(t) = (x,y)^t$ is the particle position vector, and the superscript $^o$ refers to the initial or previous condition.
The particle translational and rotational velocities vector $\mathbf{U} = (V_x\, V_y\, V_z\, \Omega_x\, \Omega_y\, \Omega_z)^t$ is calculated from a mobility matrix formulation of the hydrodynamics problem \cite{MobMatrixHammer:1996, DuffadarDavis:2007},
\begin{equation}
\mathbf{U} = \mathbf{M}\,\mathbf{F}\,,
\label{eqn:MobMatrix}
\end{equation}
where $\mathbf{F} = (F_x\,F_y\, F_z\, T_x\, T_y\, T_z)^t$ is the vector of all external forces and torques acting on the spherical particle, and $\mathbf{M}$ is a 6 x 6 matrix comprising hydrodynamic functions \cite{CoxBrennerI:1967, CoxBrennerII:1967, BrennerMobM:1961, Jeffery:1915, DuffadarDavis:2007} that model the fluid's resistance to particle motion, and that depend solely on the particle's size, fluid viscosity and particle-interface separation distance. 
The last term in the rhs in Eq. (\ref{eqn:LangevinEqn}) is included to model Brownian motion effects, and such Brownian displacements are defined as \cite{KempsBhattacharjee:2009, bacteriaTransportNelsonGinn:2005, ElimelechEtAlbook:1995} 
\begin{equation}
\mathbf{R}_\text{Br} =  \left( \begin{array}{c}
R_{\text{Br}_x}\\
R_{\text{Br}_y} \end{array} \right) = \left( \begin{array}{c}
R_{\text{Br}_\parallel} \\ R_{\text{Br}_\perp} \end{array} \right) = \sqrt{2 D_\infty dt} \left(\begin{array}{c}
\sqrt{f_4} \\ \sqrt{f_1} \end{array} \right)\mathbf{\hat{\lambda}},
\label{eqn:BrDisp}  
\end{equation}
where the vector $\mathbf{\hat{\lambda}}$ contains random numbers of the normal standard distribution, $dt$ is the time step, $D_\infty = (k_\text{B}T)/(6 \pi \mu a)$ is the Stokes-Einstein diffusivity, and $\mu$ the fluid viscosity. The functions $f_1$, $f_4$ are universal correction functions (UCFs) that account for hydrodynamic effects and depend solely on the minimum particle-surface separation distance \cite{KempsBhattacharjee:2009}. 
The time step used in all the results that follow is $dt = 1\e{-5}\,$sec.

\begin{figure*}[ht!]
\centering
\includegraphics[width=6in]{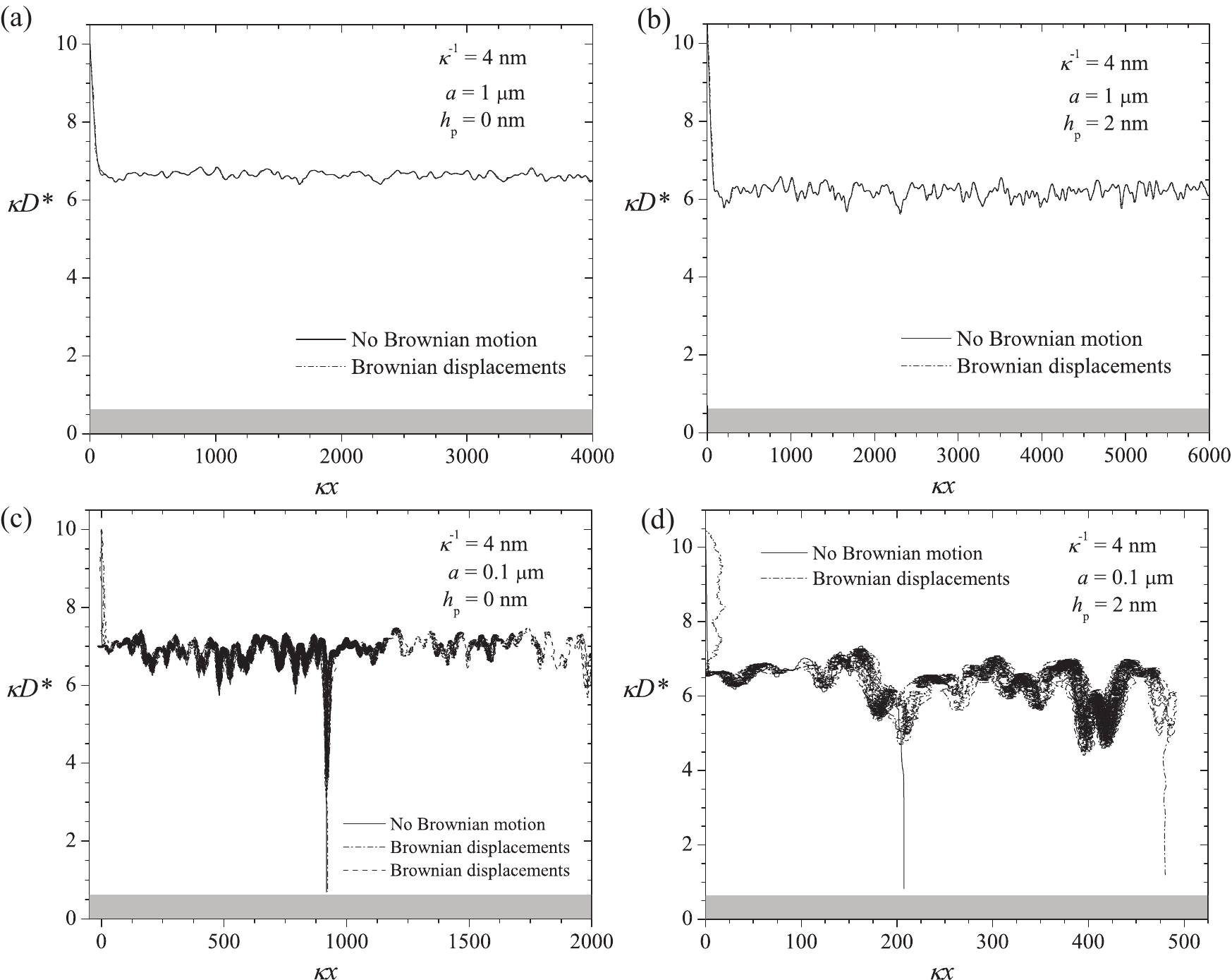}
\caption{Trajectories of particles interacting in shear flow with surfaces patterned with randomly located flat patches or cylindrical pillars with a surface area coverage of $\Theta = 0.12$. Other simulation parameters are: $\dot{\gamma} = 25$ sec$^{-1}$, A$_H$ = 5 \e{-21} J, $\psi_\text{sphere} = \psi_\text{coll} = -25$ mV, $\psi_\text{patch} = 50$ mV.
In each plot, trajectories are obtained by either neglecting Brownian effects or by adding Brownian displacements. (a) 1 $\,\mu$m radius particles interacting with patchy surfaces (h$_\text{p}$ = 0 nm). (b) 1 $\,\mu$m radius particles interacting with pillared surfaces (h$_\text{p}$ = 2 nm). (c) 0.1 $\,\mu$m radius particles interacting with patchy surfaces (h$_\text{p}$ = 0 nm). Two trajectories are shown for the case of Brownian displacements (dashed and dash-dotted lines). (d) 0.1 $\,\mu$m radius particles interacting with pillared surfaces (h$_\text{p}$ = 2 nm).}
\label{fig:HetSurf}
\end{figure*}

\section{Effects of particle size and pillar height}

The influence of Brownian motion on particle dynamics is expected to vary most significantly with system properties such as particle size and collector topography. Trajectories of small ($a = 0.1\,\mu$m) and large ($a = 1\,\mu$m) particles flowing over patchy ($h_\text{p} = 0\,$nm) and pillared ($h_\text{p} = 2\,$nm) collecting surfaces are shown in Fig. \ref{fig:HetSurf}, as plots of the dimensionless separation distance $\kappa D^*$ vs the horizontal dimensionless displacement $\kappa x$. 
$D^*$ denotes the minimum particle-flat collector separation distance and is defined as $D^* = D + h_\text{p}$. For both pillared and patchy collectors, the mobility matrix hydrodynamic functions in Eq. (\ref{eqn:MobMatrix}) were computed at the separation distance $h^* = a + D^*$, while the UCFs $f1$ and $f4$ in Eq. (\ref{eqn:BrDisp}) were calculated at the separation distance $h^* = D^*$. Examples of particle trajectories predicted with and without the effects of Brownian motion are presented in Fig. \ref{fig:HetSurf}.

In the cases of small particles interacting with both patchy and pillared collectors (Figs. \ref{fig:HetSurf}(c)-(d)), Brownian motion clearly affects particle trajectories by adding significant variations in the horizontal displacements of the particle. The spatial fluctuations observed in the dynamic profiles of large particles interacting with patchy and pillared collectors (Figs. \ref{fig:HetSurf}(a)-(b)) are due to interactions between the flowing particle and both the attractive and repulsive collector surface elements.
The attractive interactions are not strong enough to capture the particle on the surface, and the particle continues to flow until it reaches the edge of the simulated collector, given by a horizontal distance of 24.2 $\mu$m.
Interactions of small particles flowing over patchy and pillared collectors (Figs. \ref{fig:HetSurf}(c)-(d)), however, depict in all cases trajectories of particles that adhere on the collector.
For the case of a small particle interacting with a patchy collector (Fig. \ref{fig:HetSurf}(c)), however, the Brownian displacements approach can also yield trajectories of non-adhering particles, which, instead, continue to flow and translate over the entire simulated collector, represented by a horizontal displacement of 24.2 $\mu$m.

\begin{figure}
\centering
\includegraphics[width = 3.5in]{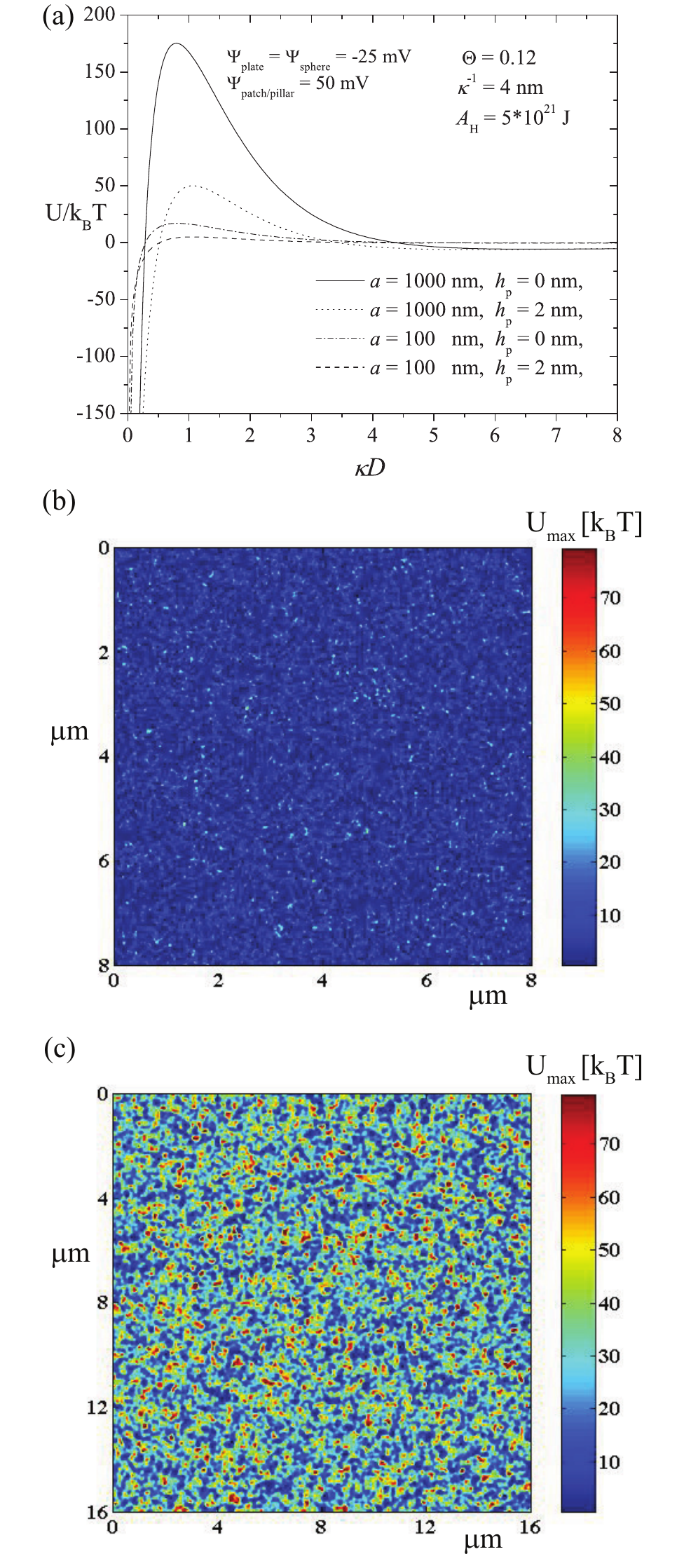}
\caption{(a) Total DLVO energy of interaction computed with the GSI$_\text{UA}$\cite{BenderskyDavis:2011} technique computed for large ($a = 1\,\mu$m) and small ($a = 0.1\,\mu$m) particles interacting with collectors patterned with flat patches ($h_\text{p} = 0$\,nm) or cylindrical pillars ($h_\text{p} = 5$\,nm) with a surface area coverage of $\Theta = 0.12$.
(b)-(c) Energy profiles for $2a$ = 200\,nm (b) and $2a$ = 2000\,nm (c) diameter particles interacting with a patchy collector ($h_\text{p} = 0$\,nm, $\Theta = 0.22$). The fractions of the collector that are favorable for adhesion (i.e. $U < k_\text{B}T$) are 32\% for the 200\,nm (b) diameter particle and 2.8\% in the case of the 2000\,nm (c) diameter particle.}
\label{fig:Energies}
\end{figure}

The total DLVO energy of interaction for large ($a = 1\,\mu$m) and small ($a = 0.1\,\mu$m) particles interacting with collectors patterned with flat patches ($h_\text{p} = 0$\,nm) or cylindrical pillars ($h_\text{p} = 2$\,nm) at a fixed surface area coverage of $\Theta = 0.12$ is computed with the 
GSI-energy averaging technique \cite{BenderskyDavis:2011}, and results are presented in Fig. \ref{fig:Energies}(a).
Within the GSI-energy averaging technique (GSI$_\text{UA}$), particle interactions with two homogeneous collectors are computed independently (for example, by implementing the GSI technique \cite{DuffadarDavis:2007, DuffadarDavisFriction:2008}) and then linearly combined using the average surface area coverage $\Theta$ as the weighting factor.   
The two homogeneous collectors are uniformly charged substrates; one of them bears the heterogeneity potential, while the other one is charged with the potential of the unpatterned areas of the heterogeneous collector.

As shown in Fig. \ref{fig:Energies}(a), the total energy of interaction for large ($a = 1000$ nm) particles interacting with patchy ($h_\text{p} = 0$ nm) and pillared ($h_\text{p} = 2$ nm) collectors presents high energy barriers, of about 175 $k_\text{B}T$ and 50 $k_\text{B}T$, respectively. 
The small ($a = 100$ nm) particles' energy profiles, however, present much lower energy barriers, of about 5 $k_\text{B}T$ for the case of the pillared ($h_\text{p} = 2$ nm) collector and about 17 $k_\text{B}T$ for the case of the patchy collector ($h_\text{p} = 0$ nm).
As previously noted, in the case of a small particle interacting with a patchy collector (Fig. \ref{fig:HetSurf}(c)), trajectories obtained with Brownian displacements 
for a surface coverage of $\Theta = 0.12$ either indicate that the particle flows until it reaches the collector's edge or show particle adhesion.
Results obtained when Brownian motion is neglected only yield trajectories that denote particle adhesion.
For the case of a small particle interacting with a patchy collector, therefore, Brownian effects become more prominent and influence the particle's tendency to adhere on the collector as a direct consequence of the intermediate magnitude of the energy barrier, of about 17 $k_\text{B}T$.

Heterogeneous collectors can be characterized by energy contour plots, as those shown in Figs. \ref{fig:Energies}(b)-(c), for $2a = 200$ nm and $2a = 2\,\mu$m diameter particles, respectively. 
An heterogeneous collector is sampled such that,
at each collector areal element, the DLVO energy-distance profile is computed with the GSI technique when the particle's center projection on the xz-plane is positioned at the center of the discrete collector element.
Thus, for each collecting surface element at which the particle center is located, a different energy-distance profile is obtained due to interactions with different regions of the heterogeneous surface. 
The magnitude of the energy barrier (U$_\text{MAX}$) is stored for each heterogeneous surface element, and these values are color-scaled to yield an energy-contour plot that highlights the various attractive and repulsive regions within the collector. Surface elements for which the energy barrier is either non-existent or negative were assigned a zero-value of the energy barrier. 
A dark blue color denotes the lowest values of the energy barrier (more attractive regions) while the regions with the highest values of the energy barrier (more repulsive areas) are marked with a dark red color. Intermediate values are scaled with other colors, as indicated in the color scale in Figs. \ref{fig:Energies}(b)-(c).

To isolate the specific influence of Brownian motion on the small particle's ($2a = 200$ nm) tendency to adhere on the patchy ($h_\text{p}$ = 0 nm) collector, we perform a few thousand simulations of particle trajectories, for a few values of the collector heterogeneity coverage $\Theta$.
At a fixed value of $\Theta$, all trajectories are simulated with the same collector surface, and in all cases, the particles are released from the same initial height of $D$ = 40 nm.
The shear flow rate is $\dot{\gamma} = $25 sec$^{-1}$ and the colloidal interactions are defined by the Hamaker constant $A_\text{H} = 5\e{-21}$ J and the Debye screening length $\kappa^{-1} = 4\,$nm.

From the direct simulation of particle trajectories, the probability of particle adhesion due to Brownian motion effects $P_\text{Br}$ is defined as
\begin{equation}
P_\text{Br} = \frac{N_D}{N_\text{tot}},
\label{eqn:ProbAdhBr}
\end{equation}
where $N_\text{D}$ is the total number of particles deposited on the patchy collecting surface and $N_{\text{tot}}$ is the total number of particles released.

Though defined in the same way as the particle collection efficiency computed in recent studies \cite{KempsBhattacharjee:2009}, the particle adhesion probability defined in Eq. (\ref{eqn:ProbAdhBr}) differs qualitatively from the collection efficiency. 
For a fixed value of the surface area coverage $\Theta$, $P_\text{Br}$ is obtained from trajectories of particles translating over \emph{one} collecting surface, and the many trajectories differ only in the random numbers that characterize the Brownian displacements. 
At each adhesion attempt, \emph{one} particle is released and allowed to interact with the heterogeneous collector in shear flow. 
The simulation runs are independent, such that the collector is patterned with randomly distributed cationic patches only, and previously adhered colloidal particles are not considered.

For the case of a small particle ($2a = 200$\,nm) interacting with a patchy ($h_\text{p} = 0$\,nm) collector, $P_\text{Br}$ values were computed for a range of values of $\Theta$, $ 0.10 \le \Theta \le 0.15$.
Each probability value was then used to construct reasonably small confidence intervals, which were computed with the Wilson score method \cite{Wilson:1927}. 
The specific number of trajectory simulations for each $P_\text{Br}$ data point therefore varied between $N_{\text{tot}}$ = 6,200 and $N_{\text{tot}}$ = 10,000, such that the difference between either end of the interval and the interval center is lower than 1.5\%.

The spread of the Brownian motion-induced adhesion probabilities can be quantified by the difference
\begin{equation}
\Delta \Theta = \Theta_\text{TL} - \Theta_\text{C}\,, 
\end{equation}
where 
\begin{equation}
\Theta_\text{TL} = \Theta(P_{\text{Br}_i} = 1),\,\;\; \forall\, \,i  \in [1, N_\text{tot}],
\end{equation}
and
\begin{equation}
\Theta_\text{C} = \Theta(P_{\text{Br}_i} = 1),\,\;\; \exists\,\,  i \in [1,N_\text{tot}].
\end{equation}
$\Theta_\text{TL}$ and $\Theta_\text{C}$ are the surface area heterogeneity coverages at which the adhesion probability $P_\text{Br}$ is equal to 1, for either all of the $N_\text{tot}$ particles released, or for at least one of them, respectively.

For a 200-nm diameter particle, the computed spread of the Brownian adhesion probability is
\begin{equation}
\Delta \Theta_\text{BD,\,2a = 200\,nm} = 0.15 - 0.11 = 0.04 
\end{equation}
where the subscripts BD denote ``Brownian displacements".

Particle trajectories that include Brownian displacements present spatial fluctuations in the flow direction, in contrast to the trajectories simulated without Brownian motion effects. These latter trajectories do not show any variance at all, such that either none of all 10,000 particles adhere on the surface at a surface coverage of $\Theta = 0.10$ or all 10,000 particles adhere on the surface at $\Theta = 0.11$ (and larger values).
For trajectories computed with Brownian displacements, however, the spatial fluctuations in the direction of flow can become meaningful enough so as to yield smaller adhesion probabilities $P_\text{Br}$ for larger values of the surface coverage $\Theta$.
The binding capability of the collector, therefore, appears to be governed not only by the total amount of attractive heterogeneity but also by the magnitude of the spatial variations in the flow direction.
Flowing particles can be translated due to Brownian displacements into more attractive or more repulsive areas of the collector, and irrespectively of the total amount of heterogeneity that patterns the collector surface.

It should be noted that these meaningful spatial variations observed in trajectories computed with Brownian displacements could also explain the unexpected finding of particles flowing at large shear rates ($\dot{\gamma} = 100$\,sec$^{-1}$) that adhere on the collector 
after translating shorter distances than those traveled by particles flowing at smaller shear rates (results not shown). 

The respective spread of the adhesion probabilities for large particles ($2a = 2\,\mu$m) is equal to zero, 
\begin{equation}
\Delta \Theta_{\text{BD,}{\,2a = 2\,\mu\text{m}}} = 0.21 - 0.21 = 0\,.  
\label{eqn:SpreadLargeA}
\end{equation} 
A zero probability spread indicates that the probability of adhesion on a specific heterogeneous collector characterized by 
a given surface coverage $\Theta$ is either 0 or 1, in complete agreement with trajectory computations 
in which the effects of Brownian motion are not considered. The computed spread in Eq. (\ref{eqn:SpreadLargeA}) thus suggests that Brownian motion has a negligible effect on the trajectories of large particles 
translating over collectors patterned with nano-scale heterogeneity.

\begin{figure}
\centering
\includegraphics{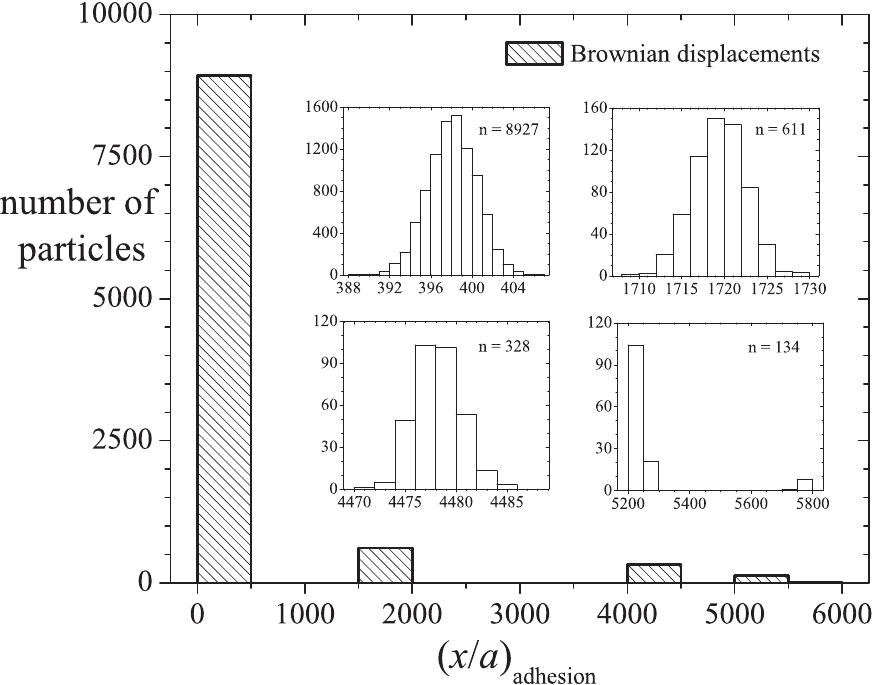}
\caption{Statistical distribution of the locations of adhered particles obtained from simulations of particle trajectories that include Brownian displacements, for a collector patterned with a heterogeneity coverage of $\Theta = 0.15$. Other simulation parameters are: $2a = 200$ nm, h$_p = 0$ nm, $\kappa^{-1} = 4$ nm, A$_H$ = 5 \e{-21} J, $\psi_\text{sphere} = \psi_\text{coll} = -25$ mV, $\psi_\text{patch} = 50$ mV.}
\label{fig:hists}
\end{figure}

For a fixed surface coverage of $\Theta = 0.15$, a histogram of the horizontal displacements the particles translate in the direction parallel to the collector before adhering (when Brownian displacements are included) is shown in Fig. \ref{fig:hists}.  
The trajectories predict adhesion not only at distances x/a $\simeq$ 398 but also in the vicinity of x/a = 1719, 4478, 5250 and x/a = 5800, suggesting the existence of other ``hot spots" \cite{DuffadarDavisSantore:2009} for particle adhesion in those regions of the collector.

\begin{figure}
\centering
\includegraphics{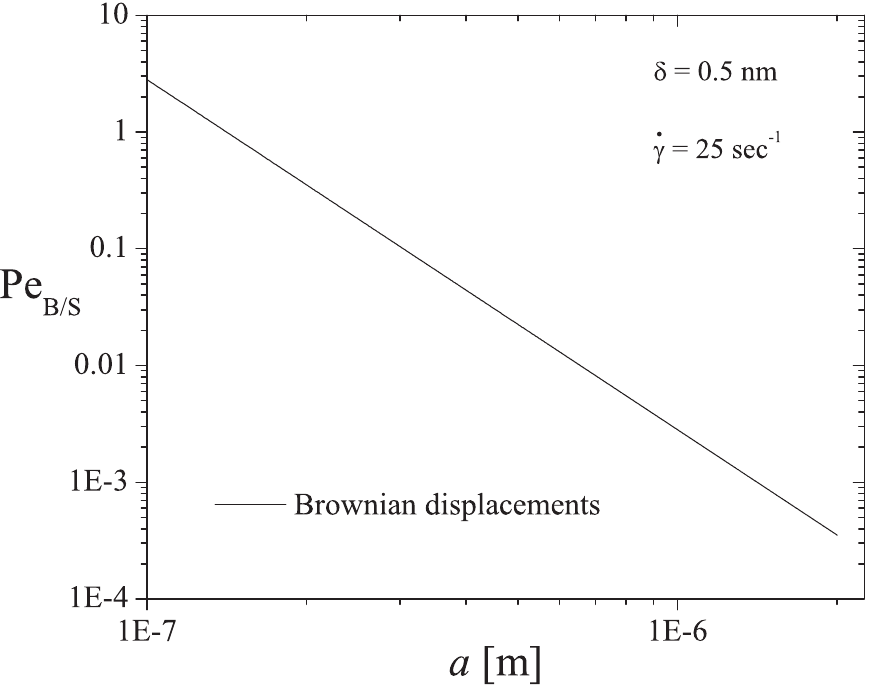}
\caption{Pe$_\text{B/S}$ numbers plotted as a function of the particle size.}
\label{fig:PeNums}
\end{figure}

\section{P\'{e}clet numbers}
In this section, we define a P\'{e}clet number that provides quantitative insight on the relative importance of the shear and Brownian forces acting in the direction of flow as
\begin{equation}
\text{Pe}_\text{(B/S)} = \frac{D_\parallel (H,a)}{\dot{\gamma} a^2},
\label{eqn:PeBrD}
\end{equation}
where $D_\parallel = f_4\,D_\infty$ is the diffusion coefficient in the direction parallel to the collector and depends solely on the particle size and distance $H$ between the particle's center and the collector \cite{ElimelechEtAlbook:1995, KempsBhattacharjee:2009}.
To avoid the unrealistic case of the particle being in physical contact with the collector, an arbitrarily small distance of $\delta = 0.5$ nm is chosen as the distance at contact, such that $H = a + \delta$. 

Eq. (\ref{eqn:PeBrD}) resembles the P\'{e}clet number defined by Kemps and Bhattacharjee \cite{KempsBhattacharjee:2009} to describe the transport of colloidal particles in the proximity of surfaces patterned with spherical asperities. In that study, however, the P\'{e}clet number indicated a greater relative importance of the shear flow with respect to the particle diffusion, such that the  diffusivity considered in the P\'{e}clet number definition was that of the bulk, or, the Stokes-Einstein diffusivity, and hydrodynamic effects were incorporated only in the computation of particle trajectories. 

In the current work, the effects of Brownian motion on particles interacting with heterogeneous collectors are studied for small separation distances in particular, with relevance to the evaluation of Brownian effects on adhesion thresholds. It thus seems appropriate to consider the hydrodynamic effects that effectively reduce the particle diffusivity as the particle approaches the collector, and to define the P\'{e}clet number in terms of $D_\parallel$ instead of $D_\infty$.

As shown in Fig. \ref{fig:PeNums}, the Pe$_\text{(B/S)}$ numbers decrease exponentially with particle size, suggesting that Brownian motion dominates over shear forces only for small particles ($a \le 200$ nm) in low-shear rate flow. Moreover, for particle radii as small as $a = 500$ nm and larger (usually used in experimental set-ups), the Pe$_\text{B/S}$ numbers fall below 0.05, suggesting that Brownian motion effects become negligible.

\begin{figure}
\centering
\includegraphics{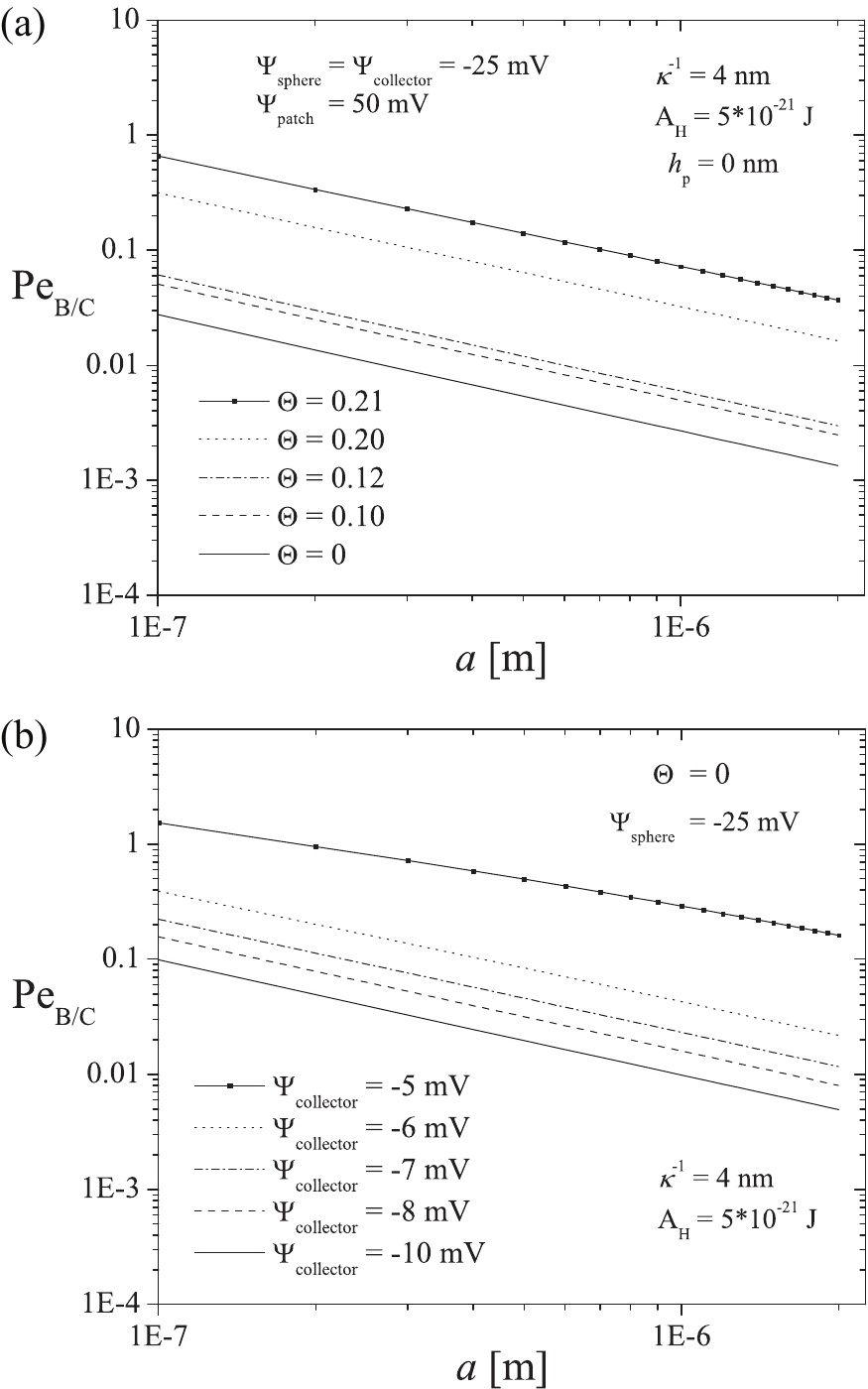}
\caption{P\'{e}clet$_\text{B/C}$ numbers as a function of the particle size, for heterogeneous and homogeneous collectors. (a) P\'{e}clet$_\text{B/C}$ numbers for a range of surface loadings $\Theta$, in the case of heterogeneous collectors. (b) P\'{e}clet$_\text{B/C}$ numbers computed for various collector surface potentials $\Psi_\text{collector}$, for uniformly charged, flat collectors.}
\label{fig:PeBrCollUtotAtDeb}
\end{figure}

To quantify the relative importance of Brownian motion effects and DLVO interactions a P\'{e}clet number Pe$_\text{B/C}$ is defined as
\begin{equation}
\text{Pe}_\text{B/C} = \frac{k_\text{B}T}{U_{\text{GSI}_\text{UA}}(D = \kappa^{-1})}\,,
\label{eqn:PeBC}
\end{equation}
where $U_{\text{GSI}_\text{UA}}(D = \kappa^{-1})$ denotes the average particle-flat collector total energy of interaction computed with the GSI$_\text{UA}$ technique \cite{BenderskyDavis:2011} at a fixed separation distance $D = \kappa^{-1}$.

P\'{e}clet numbers obtained from Eq. (\ref{eqn:PeBC}) are presented as a function of the particle size in Fig. \ref{fig:PeBrCollUtotAtDeb}, for a few values of the surface coverage $\Theta$. 

For particle sizes that are relevant for experimental purposes and a range of surface loadings for which the collector is net repulsive, 
particle-collector dynamics are dominated by DLVO interactions, rather than Brownian motion effects, as suggested by the numbers Pe$_\text{B/C} < $ 1 shown in Fig. \ref{fig:PeBrCollUtotAtDeb}(a).  
The random distribution of patches on the collector surface creates both locally attractive and locally repulsive areas, such as those showed in Fig. \ref{fig:Energies}(b) for 
a flat collector with a surface area heterogeneity coverage of $\Theta = 0.12$ sampled by a 
$2a = 400$ nm diameter particle.
Brownian motion cannot overcome the high energy barriers that characterize the locally repulsive areas, such that particle deposition on patchy collectors is not due to Brownian motion, but instead, controlled by spatially-varying DLVO interactions.

Particle deposition on slightly net-repulsive homogeneous collectors, however, can be attributed to Brownian motion effects since these systems present energy barriers that are lower than those in the locally repulsive areas within the heterogeneous collectors.
Pe$_\text{B/C}$ numbers for the case of interactions with electrostatically and topographically uniform collectors, obtained from Eq. (\ref{eqn:PeBC}) for a range of particle sizes and varying collector potentials are shown in Fig. \ref{fig:PeBrCollUtotAtDeb}(b). 
For small particles ($a \le 250$ nm) and slightly repulsive uniform collectors ($\Psi_\text{collector} = -5$ mV), Pe$_\text{B/C} > 1$ suggest that Brownian motion effects are indeed more significant than, or comparable to, DLVO interactions, and can therefore be responsible for increased deposition rates.

\section{Conclusion}

The influence of Brownian motion on the dynamics of large and small colloidal particles flowing over patchy and pillared collectors was studied by modeling particle trajectories that included the effects of Brownian displacements, in addition to those of shear and colloidal forces.

Brownian motion is expected to be meaningful only for particle-collector systems characterized by a relatively low energy barrier in the energy-distance profile. 
Computations of average energy-distance profiles obtained for interactions with heterogeneous collectors revealed that large particle interactions with both patchy and pillared collectors are limited by high energy barriers, which prevent particle deposition. 
Small particles, however, are strongly attracted to pillared collectors, such that particle adhesion is controlled by DLVO interactions and not by Brownian motion. 
Brownian motion effects, therefore, could be significant only in the case of small particles flowing over patchy collectors, 
because these systems present intermediate, or relatively-low energy barriers.

A P\'{e}clet number defined to quantify the relative importance of shear and Brownian motion effects is found to decrease exponentially with particle size. For the low shear rate flow considered, Brownian motion is seen to become more significant than, or comparable to, the shear motion only for particle radii of up to $a \simeq 200$ nm. For larger particles, Brownian motion effects are negligible, as evidenced by Pe numbers that fall below 1. To quantify the relative importance of Brownian and colloidal (DLVO) effects, Pe$_\text{B/C}$ numbers were defined as the ratio of the thermal energy to the average particle-collector energy of interaction at a fixed separation distance $D = \kappa^{-1}$. Pe$_\text{B/C}$ numbers computed for varying particle sizes and for surface loadings $\Theta$ for which average interactions are repulsive, decrease exponentially with particle size and are smaller than 1 for all the parameter ranges considered. Therefore, in the case of particles flowing over collectors patterned with nano-scale heterogeneity, Brownian motion effects are shown to be negligible also with respect to DLVO interactions.

The results presented within this work thus reveal that Brownian motion has a negligible influence on particle trajectories over collectors patterned with nano-scale heterogeneity, the non-uniform distribution of which creates locally attractive and repulsive areas within the collector. 
High energy barriers in strong locally repulsive areas cannot be overcome by Brownian motion and particle deposition on patchy collectors is controlled by spatially varying DLVO interactions. Even though Brownian motion becomes more significant than, or comparable to, shear forces for small particles in low shear rate flows, the simulations indicate that the overall adhesive behavior of the system remains unaffected by Brownian motion. In experimental studies of particle deposition on collectors covered with nano-scale heterogeneity, particle sizes are usually in the range $a = 0.5, 1 \,\mu$m. It is thus reasonable to model these systems without including Brownian motion effects.

\bibliographystyle{unsrt}

%\bibliography{beMarina}

\begin{thebibliography}{10}

\bibitem{AdamczykPatterned:2009}
Z.~Adamczyk, M.~Nattich, and J.~Barbasz.
\newblock {Deposition of colloid particles at heterogeneous and patterned
  surfaces}.
\newblock {\em Adv. Colloid Interface Sci.}, 147-148:2--17, 2009.

\bibitem{Adamczyk:2001}
Z.~Adamczyk, B.~Siwek, and E.~Musial.
\newblock {Kinetics of Colloid Particle Adsorption at Heterogeneous Surfaces}.
\newblock {\em Langmuir}, 17:4529--4533, 2001.

\bibitem{KimZydney:2004}
{M} Kim and A.~L. Zydney.
\newblock Effect of electrostatic, hydrodynamic, and {B}rownian forces on
  particle trajectories and sieving in normal flow filtration.
\newblock {\em J. Colloid Interface Sci.}, 269, 2004.

\bibitem{SandEtAl:2009}
A.~Sand, M.~Toivakka, and T.~Hjeltb.
\newblock {Influence of colloidal interactions on pigment coating layer
  structure formation}.
\newblock {\em J. Colloid Interface Sci.}, 332:394--401, 2009.

\bibitem{biosensorsKatz:2001}
E.~Katz, A.~F. B{\"{u}}ckmann, and I.~Willner.
\newblock {Self-Powered Enzyme-Based Biosensors}.
\newblock {\em J. Am. Chem. Soc.}, 123:10752--10753, 2001.

\bibitem{BiotechProducts:2007}
J.~Hubbuch and M.~R. Kula.
\newblock {Isolation and Purification of Biotechnological Products}.
\newblock {\em J. Non-Equilib. Thermodyn.}, 32:99--127, 2007.

\bibitem{DuffadarDavisSantore:2009}
R.~Duffadar, S.~Kalasin, J.~M. Davis, and M.~M. Santore.
\newblock {The impact of nanoscale chemical features on micron-scale adhesion:
  Crossover from heterogeneity-dominated to mean field behavior}.
\newblock {\em J. Colloid Interface Sci.}, 337:396--407, 2009.

\bibitem{DuffadarDavisFriction:2008}
R.~D. Duffadar and J.~M. Davis.
\newblock {Dynamic adhesion behavior of micrometer-scale particles flowing over
  patchy surfaces with nanoscale electrostatic heterogeneity}.
\newblock {\em J. Colloid Interface Sci.}, 326:18--27, 2008.

\bibitem{Kalasin:2008}
S.~Kalasin and M.~M. Santore.
\newblock {Hydrodynamic crossover in dynamic microparticle adhesion on surfaces
  of controlled nanoscale heterogeneity}.
\newblock {\em Langmuir}, 24:4435--4438, 2008.

\bibitem{SantoreKozlova:2006}
N.~Kozlova and M.~M. Santore.
\newblock {Manipulation of Micrometer-Scale Adhesion by Tuning Nanometer-Scale
  Surface Features}.
\newblock {\em Langmuir}, 22:1135--1142, 2006.

\bibitem{SantoreKozlova:2007}
M.~M. Santore and N.~Kozlova.
\newblock {Micrometer Scale Adhesion on Nanometer-Scale Patchy Surfaces:
  Adhesion Rates, Adhesion Thresholds, and Curvature-Based Selectivity}.
\newblock {\em Langmuir}, 23:4782--4791, 2007.

\bibitem{BenderskyDavis:2011}
M.~Bendersky and J.~M. Davis.
\newblock {DLVO} interaction of colloidal particles with topographically and
  chemically heterogeneous surfaces.
\newblock {\em J. Colloid Interface Sci.}, 353:87--97, 2011.

\bibitem{DuffadarDavis:2007}
R.~D. Duffadar and J.~M. Davis.
\newblock {Interaction of micrometer-scale particles with nanotextured surfaces
  in shear flow}.
\newblock {\em J. Colloid Interface Sci.}, 308:20--29, 2007.

\bibitem{AdamczykRSA:2005}
Z.~Adamczyk, K.~Jaszcz\'{o}l{}t, A.~Michna, B.~Siwek, L.~Szyk-Warszy\'{n}ska,
  and M.~Zembala.
\newblock {Irreversible adsorption of particles on heterogeneous surfaces}.
\newblock {\em Adv. Colloid Interface Sci.}, 118:25--42, 2005.

\bibitem{Walz:1998}
J.~Y. Walz.
\newblock {The effect of surface heterogeneities on colloidal forces}.
\newblock {\em Adv. Colloid Interface Sci.}, 74:119--168, 1998.

\bibitem{ModyKing:2007}
N.~A. Mody and M.~R. King.
\newblock {Influence of Brownian Motion on Blood Platelet Flow Behavior and
  Adhesive Dynamics near a Planar Wall}.
\newblock {\em Langmuir}, 23:6321--6328, 2007.

\bibitem{KempsBhattacharjee:2009}
J.~A.~L. Kemps and S.~Bhattacharjee.
\newblock {Particle Tracking Model for Colloid Transport near Planar Surfaces
  Covered with Spherical Asperities}.
\newblock {\em Langmuir}, 25(12):6887--6897, 2009.

\bibitem{UnniYang:2005}
H.~N. Unni and C.~Yang.
\newblock {Brownian dynamics simulation and experimental study of colloidal
  particle deposition in a microchannel flow}.
\newblock {\em J. Colloid Interface Sci.}, 291:28--36, 2005.

\bibitem{ErmakMcCammon:1978}
D.~L. Ermak and J.~A. McCammon.
\newblock {Brownian dynamics with hydrodynamic interactions}.
\newblock {\em J. Chem. Phys.}, 69:1352--1360, 1978.

\bibitem{AnsellDickinson:1986}
G.~C. Ansell and E.~Dickinson.
\newblock Sediment formation by {B}rownian dynamics simulation: Effect of
  colloidal and hydrodynamic interactions on the sediment structure.
\newblock {\em J. Chem. Phys.}, 85(7):4079--4086, 1986.

\bibitem{Warszynski:2000}
P.~Warszy\'{n}ski.
\newblock {Coupling of hydrodynamic and electric interactions in adsorption of
  colloidal particles}.
\newblock {\em Adv. Colloid Interface Sci.}, 84:47--142, 2000.

\bibitem{AdamczykPElayers:2006}
Z.~Adamczyk, M.~Zembala, and A.~Michna.
\newblock {Polyelectrolyte adsorption layers studied by streaming potential and
  particle deposition}.
\newblock {\em J. Colloid Interface Sci.}, 303:353--364, 2006.

\bibitem{MobMatrixHammer:1996}
{K} Chang and D.~A. Hammer.
\newblock {Influence of Direction and Type of Applied Force on the Detachment
  of Macromolecularly-Bound Particles from Surfaces}.
\newblock {\em Langmuir}, 12(9):2271--2282, 1996.

\bibitem{bacteriaTransportNelsonGinn:2005}
K.~E. Nelson and T.~R. Ginn.
\newblock {Colloid Filtration Theory and the Happel Sphere-in-Cell Model
  Revisited with Direct Numerical Simulation of Colloids}.
\newblock {\em Langmuir}, 21:2173--2184, 2005.

\bibitem{ElimelechEtAlbook:1995}
M.~Elimelech, J.~Gregory, X.~Jia, and R.~A. Williams.
\newblock {\em {Particle Deposition and Aggregation: measurement, modelling and
  simulation}}.
\newblock Colloid and surface engineering series. Butterworth-Heinemann Ltd.,
  1995.

\bibitem{CoxBrennerI:1967}
A.~J. Goldman, R.~G. Cox, and H.~Brenner.
\newblock {Slow viscous motion of a sphere parallel to a plane wall - I Motion
  through a quiescent fluid}.
\newblock {\em Chem. Eng. Sci.}, 22:637--651, 1967.

\bibitem{CoxBrennerII:1967}
A.~J. Goldman, R.~G. Cox, and H.~Brenner.
\newblock {Slow viscous motion of a sphere parallel to a plane wall - II
  Couette flow}.
\newblock {\em Chem. Eng. Sci.}, 22:653--660, 1967.

\bibitem{BrennerMobM:1961}
H.~Brenner.
\newblock {The slow motion of a sphere through a viscous fluid towards a plane
  surface}.
\newblock {\em Chem. Eng. Sci.}, 16:242--251, 1961.

\bibitem{Jeffery:1915}
G.~B. Jeffery.
\newblock {On the steady rotation of a solid of revolution in a viscous fluid}.
\newblock {\em Proc. Lond. Math. Soc.}, 14:327--338, 1915.

\bibitem{Wilson:1927}
E.~B. Wilson.
\newblock {Probable Inference, the Law of Succession, and Statistical
  Inference}.
\newblock {\em Journal of the American Statistical Association},
  22(158):209--212, 1927.

\end{thebibliography}

\end{document}